\documentclass[5p,twocolumn,times,number]{elsarticle}

\newcommand{\Dafne}       {DA\char8NE}

\newcommand{\eV}{{e\kern-.07em V}}

\def\ifm#1{\relax\ifmmode#1\else$#1$\fi}
\def\pt#1,#2,{\ifm{#1\times10^{#2}}}

\makeatletter
\newdimen\z@ \z@=0pt 
\newskip\z@skip \z@skip=0pt plus0pt minus0pt
\def\m@th{\mathsurround=\z@}
\def\ialign{\everycr{}\tabskip\z@skip\halign} 
\def\eqalign#1{\null\,\vcenter{\openup\jot\m@th
  \ialign{\strut\hfil$\displaystyle{##}$&$\displaystyle{{}##}$\hfil
    \crcr#1\crcr}}\,}
\makeatother

\def\sta#1;{\ifm{|\,#1\rangle}}    
\def\to{\ifm{\rightarrow}}

\def\bra#1;{\ifm{\langle\,#1\,|}}
\def\ket#1;{\ifm{|\,#1\,\rangle}}
\def\g{\gamma}
\def\gvir{\g^*}
\def\epiu{e^+}
\def\emeno{e^-}

\def\gg{\gamma\gamma}
\def\ee{\epiu\emeno}

\newcommand{\ba}{\begin{eqnarray}}
\newcommand{\ea}{\end{eqnarray}}
\newcommand{\be}{\begin{equation}}
\newcommand{\bnona}{\begin{eqnarray*}}
\newcommand{\enona}{\end{eqnarray*}}

\def\bbm[#1]{\mbox{\boldmath $#1$}}

\usepackage{graphics}
\usepackage{epsfig}

\usepackage{amssymb}

\journal{Nuclear Instruments and Methods in Physics Research A}

\begin{document}

\begin{frontmatter}



\title{The Low Energy Tagger for the KLOE-2 experiment}


\author[label3]{D. Babusci}
\author[label1,label2]{C. Bini}
\author[label3]{P. Ciambrone}
\author[label3]{G. Corradi}
\author[label1,label2]{A. De Santis}
\author[label1,label2]{G. De Zorzi}
\author[label1,label2]{A. Di Domenico}
\author[label1,label2]{S. Fiore\corref{cor1}}
\ead{salvatore.fiore@roma1.infn.it}
\author[label1,label2]{P. Gauzzi}
\author[label3]{M. Iannarelli}
\author[label3]{S. Miscetti}
\author[label3]{C. Paglia}
\author[label3]{D. Tagnani}
\author[label3]{E. Turri}
\address[label1]{Dipartimento di Fisica dell'universit\'a Sapienza di Roma, Roma, Italy}
\address[label2]{INFN sezione di Roma, Roma, Italy}
\address[label3]{Laboratori Nazionali di Frascati dell'INFN, Frascati, Italy}

\cortext[cor1]{Corresponding author. Address: Dipartimento di Fisica dell'universit\`a Sapienza di Roma, p.le Aldo Moro 2, 00185 Roma, Italia}

\begin{abstract}
The KLOE experiment at the upgraded DAFNE e$^+$e$^-$ collider in Frascati (KLOE-2) is going to
start a new data taking at the beginning of 2010 with its detector upgraded with
a tagging system for the identification of gamma-gamma interactions. The tagging stations for low-energy e$^+$e$^-$ will consist in two calorimeters 
placed between the beam-pipe outer support structure and the inner wall of the KLOE drift chamber. This calorimeter will be made of LYSO crystals readout by Silicon Photomultipliers, to achieve an energy resolution better than 8\% at 200 MeV.

\end{abstract}

\begin{keyword}
KLOE-2 \sep Calorimeters \sep LYSO \sep Silicon Photomultipliers.




\end{keyword}

\end{frontmatter}


\section{$\gg$ physics at KLOE}

The term ``$\gg$ physics" (or `two-photon physics") stands for the study of the 
reaction
$$
\ee\,\to\,\ee \,\gvir\gvir\,\to\,\ee \,+\, X
$$
where $X$ is some arbitrary final state allowed by conservations laws. 
Since the two-photons are in a $C = + 1$ state and the value $J = 1$ is 
excluded (Landau-Yang theorem), photon-photon scattering \cite{penn} at 
the $\ee$ colliders gives access to states with $J^{PC} = 0^{\pm +},\,2^{\pm +}$,
not directly  coupled to one photon ($J^{PC} = 1^{- -}$).  
These processes, of ${\cal O} (\alpha^4)$, with a cross section 
depending on the logarithm of the center of mass energy $E$, so that, for $E$ greater than a few GeV they dominate hadronic production at $\ee$ colliders.

%
%
%

The cross section $\sigma (\gg \to X)$ was studied at $\ee$ colliders, 
from PETRA to CESR to LEP, over the years. However, the experimental situation in the low-energy region, $m_\pi \leq 
W_{\gg} \leq 700$ MeV, \cite{cball} is unsatisfactory 
for several reasons: 
\begin{itemize}
\item large statistical and systematic uncertainties due to 
       small data samples and large background contributions;
\item very small detection efficiency and particle identification 
      ambiguities for low-mass hadronic systems.
\end{itemize}
The upgraded \Dafne~ $\Phi$ factory at the Frascati laboratories of INFN has reached a peak luminosity greater that 4x10$^{32}$ cm$^{-2}$s$^{-1}$, giving the opportunity for precision measurements of low-mass hadronic systems with a new run of the KLOE experiment (KLOE-2).

The main source of background while running on the peak of the $\phi$ resonance, i.e. $\sqrt{s}$ = 1.02 GeV, comes from $\phi$ decays, 
so that we need to
perform 
background suppression adding the information coming from a tagger system 
with an efficient detection of scattered electrons.

Most of the scattered $\ee$ are emitted in the forward directions, escaping the detection by the present KLOE detector.  
Since the energy of these electrons is below 510 MeV, they deviate from the 
equilibrium orbit during the propagation along the machine lattice. Therefore a tagging 
system should consist of one or more detectors located in well identified regions along 
the beam line, aimed to determine the energy of the scattered electrons either directly, 
or from the measurement of their displacement from the main orbit.

\section{Placement of taggers on the DA$\Phi$NE beam lattice}
In order to properly locate the  $\gg$ taggers in \Dafne, we need to accurately  track the 
off-energy particles along the machine optics. 
The code we used for optics calculations is {\tt BDSIM}~\cite{bdsim}, which allows us to estimate with similar reliability the nominal and  the off-energy particle tracks. 
With this tool we evaluated the trajectories of all particles with energy from 5 MeV up to 510 MeV. 

This simulation shows that low energy $e^+ e^-$ , i.e. the ones with energy below 250 MeV, will exit from the beam pipe within one meter from the interaction point. These particles can be identified by a detector very close to the beam pipe in the horizontal plane. Off-energy $\ee$ detectable in this way are only those traversing the beam pipe in a region where no magnetic element is present, where placing a detector is possible. Combining all these constraints, the allowed region is between the first two quadrupoles close to the interaction point. In that place, most of the off-energy $\ee$ coming out of the beam pipe will have an energy between 160 and 230 MeV and an average angle of 11 degrees with respect to the beam axis. The presence
of these particles will tag radiated photons in the main detector, with
energies between 280 and 350 MeV.
 
This simulation also shows that for these off-energy $\ee$ there is only a rough correlation between particle energy and trajectory.
For this reason is 
necessary to design an energy-sensitive detector, i.e. a calorimeter, instead of a position-sensitive one.

A further constraint to be considered is the presence of the KLOE drift chamber inner wall and the beam pipe support structure. This limits the radial dimension of the detector between 13 and 21 cm, when accounting also for the clearance necessary to install the inner detectors and the beam pipe.

The low-energy station of the $\gg$ tagger will therefore consist of two identical calorimeters, symmetrically placed at both sides of the interaction point. 
This inner detector will be referred as Low Energy Tagger or LET. 
These detectors must reach high energy and time resolutions in order to:
\begin{itemize}
\item improve the invariant mass resolution on the reconstruction of $\gg$ decay products, 
\item allow correlation between the bunch crossing and the detected events, and reject accidental particles coming from the machine background.
\end{itemize}

A second tagging station, called High Energy Tagger or HET~\cite{archilli}, will be placed after the first bending dipole, to detect high-energy $\ee$. In order to get 5\% energy resolution on the LET-LET coincidence 
we must reach 8\% energy resolution on the single LET station.
Combining all the three possible tagging combinations, we can get 
a clean data sample of $\gg$ physics events  with an efficiency of about 10\%. In the first year of KLOE-2 data taking, aiming at $\sim$ 5 fb$^{-1}$ of integrated luminosity, this will provide us an effective integrated luminosity of about 500 pb$^{-1}$ for $\gg$ physics.

\section{Technology: crystals and photosensors}

By the above-mentioned requirements we can impose some technological constraints to our detector:
\begin{itemize}
\item it must be made by high-$Z$ material, in order to place a detector with good shower containment 
in a small volume so to minimize shower leakage. Crystals are a good option: these should have a high light yield, and a fast emission time, in order to get the desired timing and energy capabilities,
\item photodetectors must be thin, with high gain and insensitive to magnetic fields.
\end{itemize}

To cope with all these different requirements, suitable choices are offered by the most recent scintillating crystals and photodetectors. To this purpose the Lead Tungstate (PbWO$_4$) and the new Cerium doped Lutetium Yttrium Orthosilicate (LYSO) crystal scintillators have been considered, coupled to  Silicon Photomultipliers (SiPM) photodetectors.

These two scintillators have a radiation length X$_0$ of $\sim$1 cm, a Moliere radius R$_M$ $\sim$2 cm, and a fast emission time of 10 to 40 ns. The two crystals differ significantly for the light yield: while LYSO has a relative output of 83\% with respect to NaI(Tl), PbWO$_4$ has hundreds of times smaller light output.
Dedicated measurements of these two crystals, exposed to cosmic rays and electron test beams, have been used to choose the best option between the two scintillators. 

SiPM (also known as G-APD or MPPC) are arrays of very small Avalanche Photodiodes (APDs) operated in Geiger mode, parallel-connected via individual quenching resistors. The sum of the ``digital'' signals of each G-APD, one for each photoelectron, makes the device analog again. SiPM response is linear until 
the number of incident photons is more than 20\% of the number of  pixels. Then  the probability for two photons to fall into the same pixel becomes non-negligible, inducing a non-linear response ({\it pixel saturation}).

Main advantages with respect to APDs and PIN diodes are:
\begin{itemize}
\item high gain (10$^6$ with respect to PIN diodes, 10$^4$ with respect to APDs),
\item no nuclear counter effect due to leaking shower particles reaching photosensor, thanks to reduced thickness,
\item almost no avalanche fluctuations from excess noise factor,
\item low bias voltage (70V or less)
\end{itemize}

\section{Front-end electronics}
The front end electronics has been custom designed to satisfy the detector requirements and to be compatible with the KLOE Electromagnetic Calorimeter (EmC) readout chain~\cite{nimcalo}.

From the detector point of view, the main requirements are
\begin{itemize}
\item a very stable, low noise power supply for the SiPMs, 
\item a working voltage setting and monitor for each SiPM channel,
\item a low noise, good linearity, low power consumption preamplifier,
\item a good packing factor.
\end{itemize}
At the same time to be compatible with the EmC readout, the preamplifier output must be 
well matched with the input stage of the KLOE SDS (Splitter-Discriminator-Shaper) board.

\begin{figure}[htb]
\begin{center}
\psfig{file=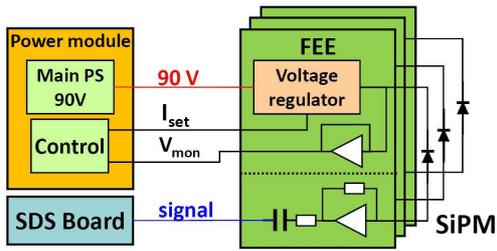,width=6.5cm}
\end{center}
\caption{Block diagram of the LET front-end electronics}
\label{fee}
\end{figure}

To fulfil the specifications, a transimpedance preamplifier with embedded voltage regulator has been developed. The block diagram of the complete system is shown in fig.~\ref{fee}.
The remote power module generates the main power supply for the on-board voltage regulator  and allows setting and  monitor of SiPM working voltage. 
A main DC voltage of 90 V with 100 mV of residual ripple 
can be distributed in parallel to several channels 
The working voltage of each SiPM is set via a programmable current sent to the correspondent  voltage regulator. The use of a control current instead of a control voltage is needed to reach the required setting precision, solving  the problem of the cables and contacts series resistance.

To obtain a very stable voltage minimizing the number of components, the  voltage regulator uses a new architecture based on a constant current source and a voltage reference source to control a variable impedance to regulate the input voltage.
The output voltage can be regulated in the range 60 - 80 V with a precision of few mV and the maximum load current is 1mA. The obtained load regulation is 0.002\% with a long term stability of 0.02\%. The RMS of output noise (10 Hz- 10 kHz)  is 10 $\mu$V and the ripple rejection ratio is $>$70 dB. The supplied current is less than 2 mA.

A two transistors configuration with bootstrap technique has been adopted for the transimpedance preamplifier. A low noise microwave transistors with a transition frequency $f_T$ of 5 GHz  and a very low base spread resistance $r_{BB'}$ have been used to match the noise requirements (2$\mu$V/$\sqrt{Hz}$). Besides, the use of only two transistors operating  at low collector current allows to reach a power dissipation of 24 mW. The bootstrap technique is adopted to  preserve  the fast rising edge of the signal with a  high detector capacitance. 

The preamplifier with the embedded voltage regulator has been packed in a 10x20 mm2 board that can be mounted directly on the face of the LYSO crystal.

\section{Detector simulation}
\begin{figure}[htb]
\begin{center}
\psfig{file=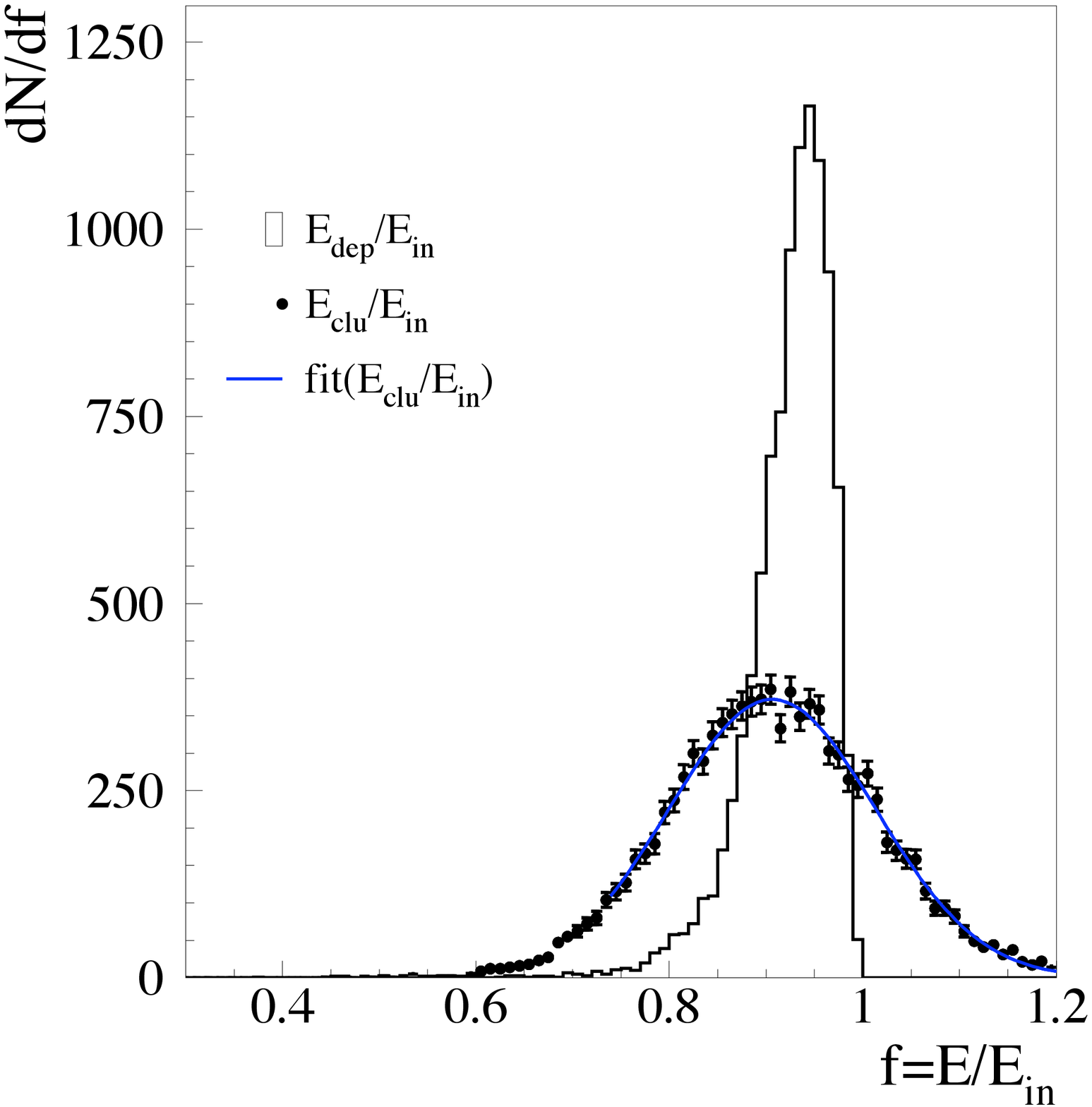,width=4.2cm}
\psfig{file=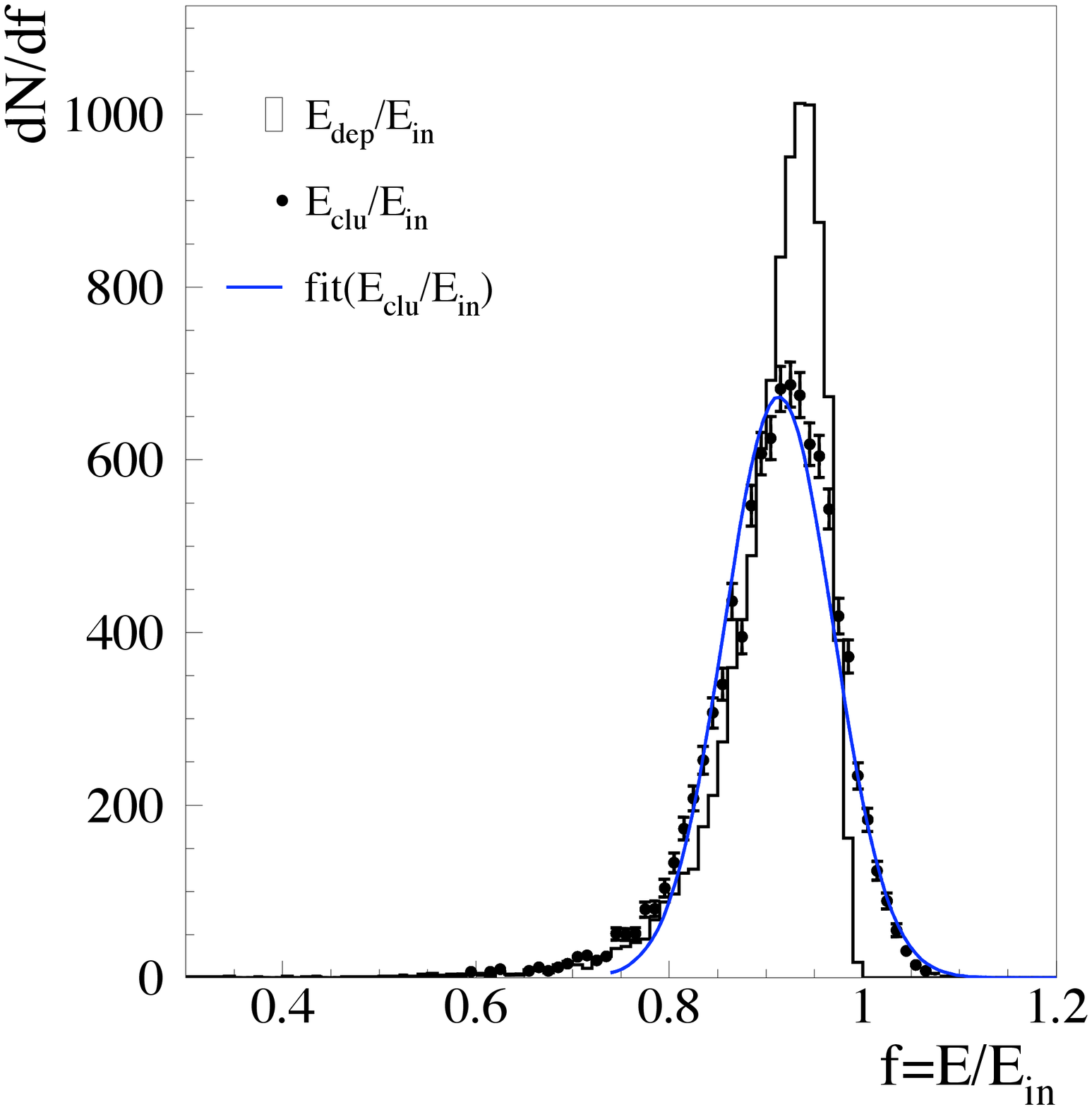,width=4.2cm}
\end{center}
\caption{Deposited and reconstructed energy in one LET detector, averaged over the impinging off-energy beam particles. Each plot shows for LYSO (left) and PbWO$_4$ (right) the fraction of deposited energy (solid) and reconstructed energy (dots) over the impinging energy. To evaluate the reconstructed energy distributions, 0.5 p.e. for PbWO$_4$ and 3.5 p.e. for LYSO have been used. Notice that the deposited energy fraction is similar in both cases and equal to 91\%.}
\label{let_mcreso}
\end{figure}

The small available space for the LET 
 makes it necessary to carefully choose the dimension of the active part of the detector  to maximize the acceptance to $e^+ e^-$ while minimizing the shower leakage. In order to do that, extensive Monte Carlo simulations, made with the GEANT4 simulation package, have been used to evaluate the expected energy and time resolution with different geometric options for the final design. Simulations of  LYSO and PbWO$_4$ crystals energy response allow us to evaluate 
the best fitting properties for the LET. A full simulation of the LET active volume and beam trajectories is used to determine the detector response to the expected off-momentum particles. By changing the simulated detector dimensions we will find the best compromise between available space and energy resolution. 
Each detector will consist of crystals pointing to the average direction of the off-energy particles, i.e. 11 degrees with respect to the z axis, centered on the horizontal plane.  Distributions of the energy deposits, for PbWO and LYSO crystals, are shown by solid line in fig.~\ref{let_mcreso} for a 60x80x130 mm$^3$ detector. 

\section{Test beam results}

An intensive test beam campaign is being carried on from May to June 2009, at the Beam Test Facility (BTF)~\cite{btf} of the Laboratori Nazionali di Frascati of INFN. Here, single electrons of energy ranging from 50 to 500 MeV are available with a rate of 50 Hz. Single crystals, as well as a crystal matrix with dimensions close to the LET final design, have been exposed to the beam in order to characterize scintillator and photodetector behaviour.

\begin{figure}[htb]
\begin{center}
\psfig{file=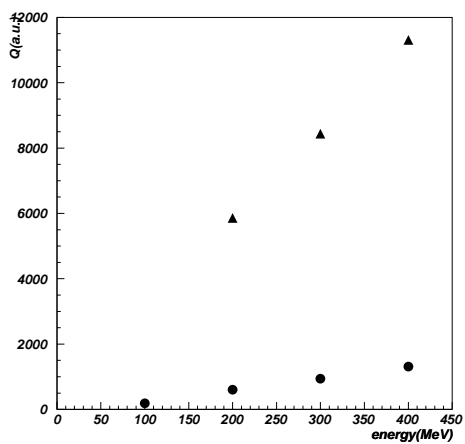,width=6.5cm}
\end{center}
\caption{Average response for single PbWO$_4$ (circles) and LYSO (triangles) crystals, as a function of the impinging electron energy.}
\label{lyso-pbwo}
\end{figure}

The first tests aimed at choosing the best option between PbWO$_4$ and LYSO: single crystals of 20x20x130 mm$^3$ were coupled to Hamamatsu MPPC 3x3 mm$^2$ S10362-33-050C, with 3600 pixels. SiPM gain was set to 1x10$^6$, and an amplifier gain equal to 20 was used in the above described front-end electronic boards. PbWO$_4$ was coupled to SiPM by means of Saint-Gobain BC-630 optical grease. No grease was used for LYSO coupling, where instead a thin yellow optical plastic filter was used to avoid signal saturation.
Crystals were exposed to energies from 100 to 400 MeV, and the collected charge is shown in fig.~\ref{lyso-pbwo} as a function of the beam energy. 
As expected, the two crystals shown very different light yields. PbWO$_4$ light output resulted in $\sim$0.2 p.e./MeV, too poor for the LET requirements, while LYSO had $\sim$2 p.e./MeV, demonstrating to be a good option to be used in this calorimeter. In both cases the SiPM worked in a non-saturated regime, which means that the number of photoelectrons was below 20\% of the total number of SiPM pixels at highest energy. 

\begin{figure}[htb]
\begin{center}
\psfig{file=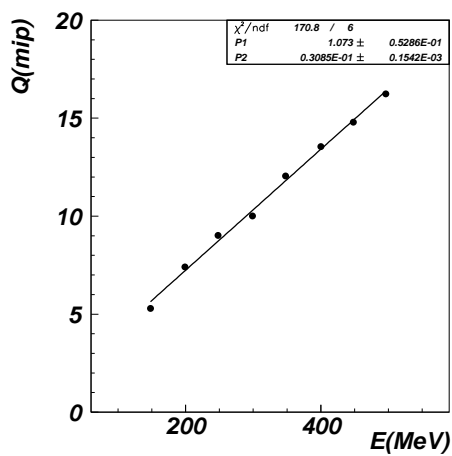,width=6.5cm}
\end{center}
\caption{Energy response in minimum-ionizing particles' units for the crystal matrix, with respect to beam energy, for LYSO crystals only}
\label{lineartb}
\end{figure}

\begin{figure}[htb]
\begin{center}
\psfig{file=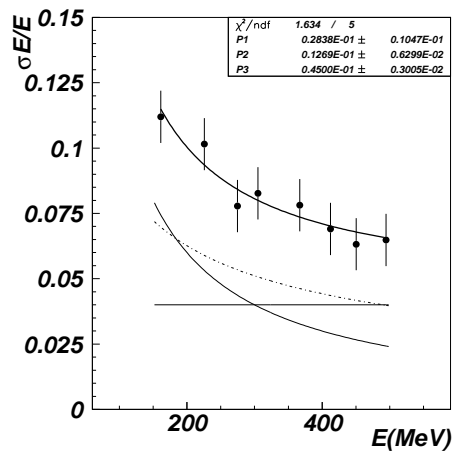,width=6.5cm}
\end{center}
\caption{Energy resolution for the crystal matrix, as a function of the beam energy, obtained by summing up LYSO crystals only. Individual contributions from stochastic term (solid), noise (dotted) and constant term (horizontal dotted) are shown below the fit data.}
\label{eneresol}
\end{figure}

Then, a crystal matrix of 60x55x130 mm$^3$ made of a core of LYSO crystals, surrounded by PbWO$_4$ ones, was exposed to the BTF electron beams with energies from 150 to 500 MeV. Each crystal was coupled to one or two SiPMs of the same type as in the tests on single crystals. No optical grease and no optical filter was used, and the LYSO crystals were wrapped into black paper tape. SiPM gain was set to 10$^5$, and the amplifier gain was set to 2 for all the SiPM placed on LYSO crystals. The above-described front-end electronics was coupled to SiPMs and to KLOE DAQ chain, to test full detector integration.
At  the moment data analysis is still in a preliminary stage, and the results presented in the following only refer to the LYSO crystals (about 2/3 of the matrix). Linearity of the matrix response has been checked by summing over the LYSO crystals for electron energies from 150 to 500 MeV. The collected charge is shown in fig.~\ref{lineartb}. 
A linear response of LYSO+SiPM from 150 up to 500 MeV is achieved. This implies that less than 10$^3$ photoelectrons are collected at the maximum energy, which corresponds to $\sim$2 photoelectrons/MeV, consistently with what expected from the hardware setup.
Preliminary results for the energy resolution are shown in fig.~\ref{eneresol}. 
The stochastic term is 2.8\%/$\sqrt{E(GeV)}$, that is consistent with the $\sim$2 photoelectrons/MeV expected. The noise term was evaluated from the pedestals of the electronic channels, and contributes with 0.7\%/E to energy resolution. Some additional noise due to test beam setup, currently under investigation, brings the total noise contribution to 1.2\%/E. A constant term of $\sim$4\% is fully dominated by leakage and will be reduced once summing over all the crystals. 

\section{Final design of the LET detectors}

Once the analysis of the test beam data will be completed, we will 
finalize the design of the LET calorimeters. 
Preliminary results suggest to use LYSO crystals, by wrapping them with Tyvek reflecting foils, which will increase the light yeld of 4 times with respect to the present test beam configuration. In order not to saturate SiPM pixels, the 3x3 mm$^2$ 14400 pixel S10362-33-025C SiPM by Hamamatsu will be used. In this way we will increase the photoelectron statistics, and benefit also from less temperature dependence for SiPM gain and dark counts.
The final calorimeter dimensions are being evaluated by technical drawings and Monte Carlo simulations, in order to optimize the energy resolution for a given geometry. We will also equip the SiPM support structure with a thermal sensor array, to monitor the temperature variations, even though these are expected to be  small in the center of the KLOE detector.
In order to frequently calibrate the calorimeters, a combined system of light pulsing and radioactive source calibration will be used for each crystal.

\bigskip
We thank B.~Buonomo, M.~Martini, F.~Happacher, and the
LNF mechanical workshop staff for their help during the assembling and
test beam phases.





\end{document}
\endinput